# Graphene Nanotechnology for the Next Generation Nonvolatile Memory

Md. Nahid Hossain and Masud H Chowdhury
Computer Science and Electrical Engineering, University of Missouri – Kansas City, Kansas City, MO 64110, USA
Email: mnhtyd@mail.umkc.edu; nahidapece@gmail.com; masud@ieee.org,masud@umkc.edu

*Abstract*— As conventional silicon technology is approaching its fundamental material and physical limits with continuous scaling, there is a growing push to look for new platform to design memory circuits for nanoelectronic applications. In this paper we explore new design concept of nonvolatile memory based on graphene nanotechnology. The investigation focuses on two forms of graphene based field effect transistor (FET) –carbon nanotube FET (CNTFET) and graphene nanoribbon FET (GNRFET). The analysis reveals that GNRFET with a high trapping capable oxide layer is suitable for ultra-high-density nonvolatile memory.

## I. INTRODUCTION

Nonvolatile memory can retain stored information even when power is turned off, and it does not need periodic refreshing like volatile memory. Nonvolatile memory includes all forms of Read-Only-Memory (ROM), such as, programmable read-only memory (PROM), erasable programmable read-only memory (EPROM), electrically erasable programmable read-only memory (EEPROM), and flash memory. Nonvolatile memory is typically used for the task of secondary storage, or long-term persistent storage. Unfortunately, non-volatile memory either costs more or performs worse than volatile random access memory. Non-volatile data storage can be categorized in electrically addressed systems (read-only memory) and mechanically addressed systems (hard disks, optical disc, magnetic tape, and holographic memory). Electrically addressed systems are expensive, but fast, whereas mechanically addressed systems have a low price per bit, but are slower.

High performance computing requires ultra-high density, low-power and high-speed memory. With technology scaling traditional CMOS technology based memory devices are experiencing severe challenges due to excessive thermal stress, power consumption, process and parametric variations, leakage and other types of noise. It is reported that carbon/graphene based field effect transistors such as carbon nanotube field effect transistors (CNTFET) and graphene nanoribbon field effect transistor (GNRFET) can change the conductivity by a factor of 1000 or more. It is also anticipated that the CNT and GNR could minimize the thermal overhead due to their high thermal conductivity [3].

Recently, there has been a surge of interest in exploring CNT and GNR based devices. Graphene based memory is a critical component of this new direction of research. Nano-electro-mechanical (NEM) switches [7], CNT based nanorelay [8], CNT based oxide–nitride–oxide film and nanoscale channel [6], CNT based non-volatile Random Access Memory (RAM) for Molecular Computing [9] are some examples of recent investigations and proposals regarding the next generation nonvolatile memory.

## II. PRESENT SCENARIO

The demand for nonvolatile semiconductor memory, particularly flash memory, has been growing explosively and further growth is expected. Over the last two decades, improvement of nonvolatile memory technology has been one of the most important factors behind the remarkable growth of hand-held and mobile devices, and many other consumer electronics. Cost of flash memory is related to the process complexity and silicon occupancy. And the performance of flash memory is dependent on MOSFET $I_{on}/I_{off}$ ratios, While functionality and versality of flash memory is dependent on the density of MOSFET that can be provided in a memory chip. Nonvolatile flash memory size has been going down steadily over the last two decades. During the last 20 years, flash memory cell size has been decreased by ~1/1000. Figure 1 shows the NOR flash memory scaling history and Figure 2 illustrates the same for NAND flash.

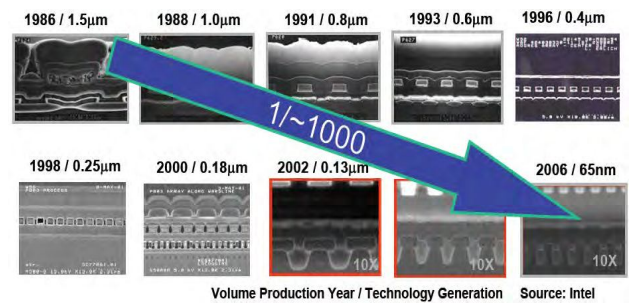

Figure 1: NOR flash memory scaling history in 20 years from 1996 to 2006 [source: Intel]

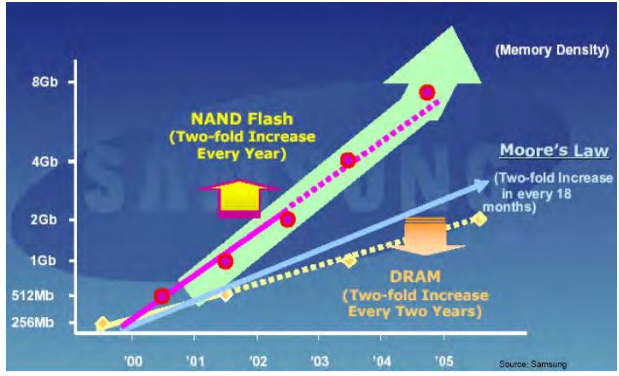

Figure 2: NAND flash memory technology scaling trend, NAND flash density increases two-fold every year [source: Samsung]

However, the conventional nonvolatile flash memory is currently facing serious challenge as industry is trying to scale it below 32nm technology. One of the major bottlenecks is the inability to scale down gate length due to severe short channel effects (SCE) and gate scaling saturation leading low programming efficiency. Cell capacitance lowering and inability to scale the gate stack due to leakage control requirements are also affecting flash memory design beyond 32nm.

Another consequence of scaling is the reduction of $I_{on}/I_{off}$ ratio in conventional MOSFET devices. In conventional short channel MOSFT $I_{on}$ is specified as the saturation current ($I_{dsat}$) for a particular set of gate and drain voltages. $I_{off}$ is the subthreshold leakage current when gate voltage is zero or below the threshold voltage. This ratio is going down due to excessive increase of leakage currents. Traditionally, only subthreshold conduction used contribute leakage, but now there more many other leakage sources that are raising the off current alarmingly.

Scaling improves cost, speed, and power per function with every new technology generation. All of these attributes have been improved by 10 to 100 million times in four decades --- an engineering achievement unmatched in human history. When it comes to ICs, small is beautiful. But, the question is what are the barriers to further scaling? Can scaling go on forever? Table 1 shows the projection of International Technology Roadmap for Semiconductor (ITRS). If we closely examine this table we can deduct that continuous $I_{off}$ will become an increasingly more critical issue for IC design unless we move to subthreshold circuit design.

Table 1: Excerpt of 2003 ITRS technology scaling from 90nm to 22nm. HP:High Performance technology. LSTP: Low Standby Power technology for portable applications. EOT: Equivalent Oxide Thickness.

| Year of Production | 2004 | 2007 | 2010 | 2013 | 2016 |
|---|---|---|---|---|---|
| Technology Node (nm) | 90 | 65 | 45 | 32 | 22 |
| HP physical Lg (nm) | 37 | 25 | 18 | 13 | 9 |
| EOT(nm) (HP/LSTP) | 1.2/2.1 | 0.9/1.6 | 0.7/1.3 | 0.6/1.1 | 0.5/1.0 |
| Vdd (HP/LSTP) | 1.2/1.2 | 1.1/1.1 | 1.1/1.0 | 1.0/0.9 | 0.9/0.8 |
| Ion/W,HP (mA/mm) | 1100 | 1510 | 1900 | 2050 | 2400 |
| Ioff/W,HP (mA/mm) | 0.05 | 0.07 | 0.1 | 0.3 | 0.5 |
| Ion/W,LSTP (mA/mm) | 440 | 510 | 760 | 880 | 860 |
| Ioff/W,LSTP (mA/mm) | 1e-5 | 1e-5 | 6e-5 | 8e-5 | 1e-4 |

### III. PROPOSED GNRFET FOR MEMORY DESIGN

The block diagram of a basic memory cell is shown in Figure 3, where in addition to the *input* and *output* ports there are two control signals (*Select* and *Read/Write*) to coordinate the operation of the memory cell. Figure 4 shows the 3D structure of a GNRFET with few layers of graphene nanoribbon channel. Source(S) and Drain (D) are connected through a multi layered graphene channel. The top gate is isolated by a high k-dielectric. Similarly, the back gate is separated by three layers which are clearly visualized in Figure 5.

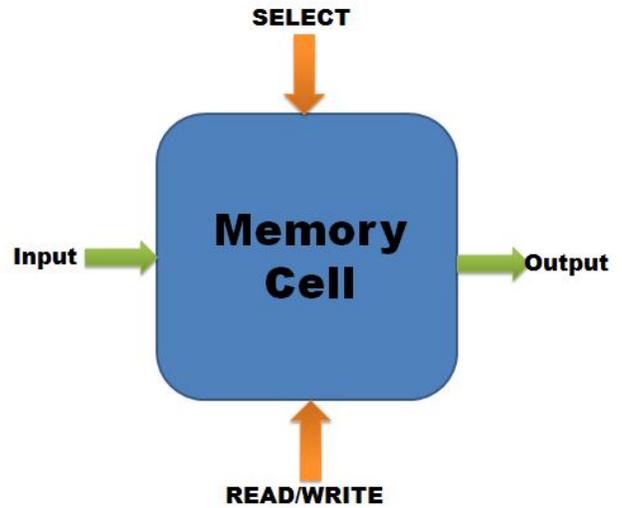

Figure 3: Conceptual view of Basic memory cell unit for Read/Write operation.

### IV. POTENTIALS OF GNRFET IN NON-VOLATILE MEMORY

*High $I_{on}/I_{off}$ ratio GNRFET memory unit*

Very high-speed memory requires higher $I_{on}/I_{off}$ ratio Graphene nanoribbon Field Effect Transistor (GNRFET). Alternatively, the $I_{on}/I_{off}$ ratio directly depends on the bandgap of Graphene nanoribbon (GNR). The experimental $I_{on}/I_{off}$ ratios reported to date, including bilayer GNRs clearly support theoretical predictions and show promise for graphene transistor based memory cell [3]. In terms of flash-type memory operation, a significant threshold voltage shift is essential for obtaining a large $I_{on}/I_{off}$ ratio [6].

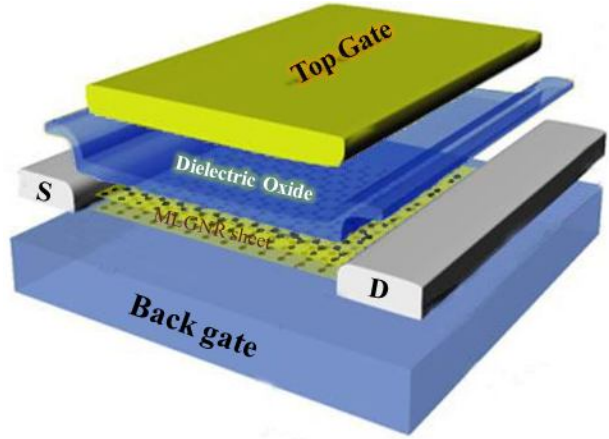

Figure 4: 3D view of Graphene nanoribbon Field Effect Transistor (GNRFET) memory cell whose source, drain, top gate and back gate act as input, output, select, read/write operations respectively.

*High mobility*

Despite the mobility reduction after the gate oxide deposition, graphene mobility clearly exceeds the universal mobility of silicon over the entire measured range when the contact resistance is accounted for. This is particularly encouraging since this result has been achieved with an evaporated gate oxide with a high charge trap density. Similar promising results have been obtained for epitaxial grapheme **[3]**.

*Hysteresis*

Hysteresis is often observed in graphene FETs as the gate voltage is swept from negative to positive direction and vice versa. This can occur despite measuring in vacuum conditions and is a strong indicator of charge traps near the graphene/ insulator interface [3].

*Low power operation*

Low-power consumption is another important parameter that affects memory performance. The GNRFET can be used in subthreshold conduction. In addition, CNT based nanotube interconnects are better solution than metal counterpart such as Cu, Al.

*Compact dimension*

Non-volatile memories are always arranged in densely compact form. For instance, an 8GB ROM 4 bits per memory cell architecture we have to accommodate 17179869184 transistors within a tiny space.

*Device Oxide*

High quality gate oxide is recommended for non-volatile memory because they have to trap substantial amount of charges. Obviously, better higher k-dielectric would be perfect choice. Additionally, gate oxide thickness should not scale down beyond an optimal limit to suppress electron tunneling.

*Low-loss interconnect and heat conduction*

Non-volatile memory blocks are too much congested; produce huge heat during operation, heat transfer rate is lower than other Integrated circuits. So, efficient, less consumption material interconnects are recommended so that it produce less heat. Hence it will perform better.

## V. MODELING

The cross section of our dual gate Graphene nanoribbon Field Effect Transistor (GNRFET) is shown in Figure 5 . Dual gate facilitate us parallel and simultaneous operation. Both read/write and select operations can be performed simultaneously. Back gate is separated from the transistor by P-type Si and $SiO_2$ while top gate is isolated by a high k-dielectric material. The few layer Graphene nanoribbons are visualized with yellow color between source and drain.

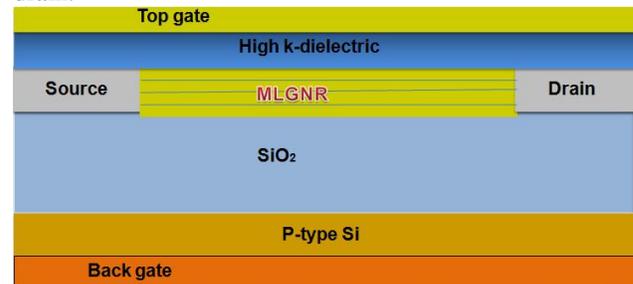

**Figure 5**: Cross sectional view of proposed Graphene nanoribbon Field Effect Transistor (GNRFET).

For low power operation, we kept the thickness of the $SiO_2$ and top gate dielectric at an optimal level so that our transistors consume less power.

The complete structure of our proposed model is shown in Figure 6. We have used few layers Graphene nanoribbon Field Effect Transistor (GNRFET) as memory cell unit and carbon nanotube (CNT) as interconnects. The dual gated Graphene nanoribbon Field Effect Transistor (GNRFET) is immune to leakage current and power dissipation. Word line (WL) are connected to each gate of the Graphene nanoribbon Field Effect Transistor (GNRFET) whereas Data lines (DL) are connected to the source or drains(source and drain are just image of each other) of the transistors through current limiting resistors. REFs or GNDs are introduced intermittently in order to suppress crosstalk. Bundles of carbon nanotubes (CNTs) are proving smooth carrier mobility, low contact resistance, extended conduction channel and easier alignment.

The $I_{on}/I_{off}$ ratio of these grapheme nanoribbon (GNR) based non-volatile memory cell unit is far greater than traditional CMOS memory cell. This ensures fast operation of the proposed non-volatile memory which promotes faster processing and execution in computing. In this model, every GNRFET perform better because REF and GND are perfectly maintained which also minimize leakage current.

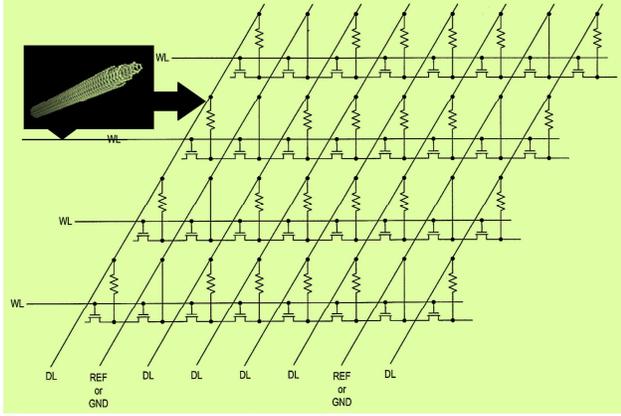

**Figure 6**: Schematic of the proposed non-volatile memory model. Here dual gate Graphene nanoribbon Field Effect Transistors (GNRFETs) is used as memory cell unit and Bundles of carbon nanotubes (CNTs) are served as Date line (DL) and Word Line (WL) interconnects.

## VI. RESULT AND DISCUSSION

*SELECT operation* is used to choose a specific memory cell. Thin layer top gate is utilized for this purpose. The $I_{ds}$-$V_{gs}$ characteristics of GNRFET are simulated in for three different drain-source voltages ($V_{ds}$). The currents $I_{ds}$ are plotted in the y-axis while the applied top gate voltages ($V_{gs}$) ranging -20mV to 100mV are plotted along x-axis. The graph reveals the following important observations. First, only mV voltages at top gate and bias voltages ($V_{ds}$) are enough to create conduction in Graphene nanoribbon Field Effect Transistor (GNRFET). The main mechanism behind this phenomenon is thin high k-dielectric at the top gate which ensures low voltage demand. This clearly indicates very low power operation in GNRFET. Secondly, Hysteresis is clearly projected in . For instance, when top gate positive voltage is decreased to zero volt, current ($I_{ds}$) should be ideally zero. But in our $I_{ds}$-$V_{gs}$ characteristics, we are always getting some current which proves that considerable quantity of charges is still now in the MLGNRFET channel. Hence, we can strongly declare that hysteresis, the most desired property for memory devices, present inherently in MLGNRFET.

*Read/write operation* is used to perform read or write operation on a specific memory cell. Comparatively thicker back gate is utilized for this purpose. The $I_{ds}$-$V_{gs}$ characteristics of GNRFET are simulated for three different drain-source voltages ($V_{ds}$). The currents $I_{ds}$ are plotted in the y-axis while the applied back gate voltages ($V_{gs}$) ranging -40V to 100V are plotted along the x-axis. The graph reveals the following significant observations. First, only mV drain-source biasing voltages ($V_{ds}$) are enough to maintain ON-state in Graphene nanoribbon Field Effect Transistor (GNRFET). This evidently indicates very low power operation in GNRFET. One thing should be noted that the back gate input voltages are greater than top input voltages because of additional distance from the GNRFET channel. Secondly, Hysteresis is also clearly projected in Figure 8. For example, when back gate positive voltage is decreased to zero volt, current ($I_{ds}$) should be perfectly zero. But in our $I_{ds}$-$V_{gs}$ characteristics, we are noticing significant amount of current which proves that some charges are still now in the MLGNRFET channel. Therefore, we can strongly show that hysteresis, the most wanted property for memory devices, present intrinsically in MLGNRFET.

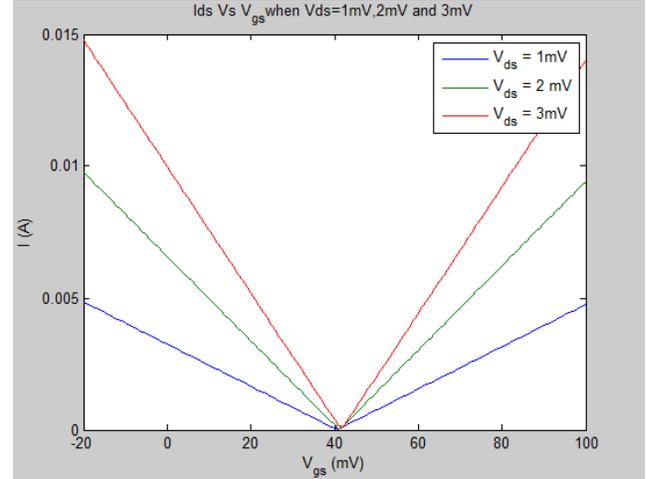

Figure 7: (Ideal case) Ids–Vgs characteristics of the designed GNRFETs with L=1.3μm, W=2.7μm. The minimum conduction point is 40 mV.

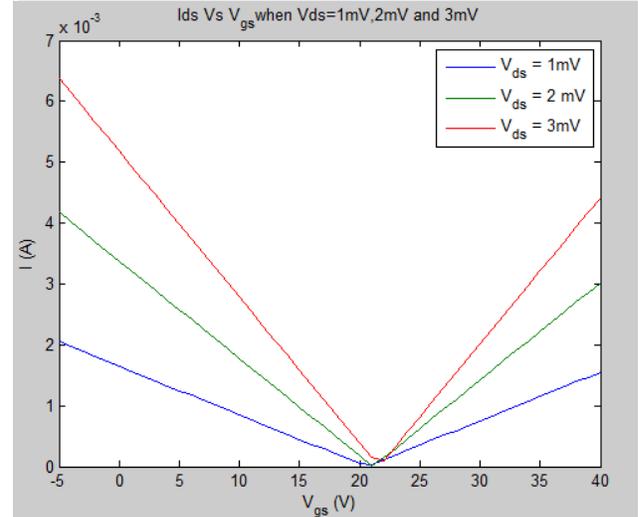

Figure 8: (Ideal case) Ids–Vgs characteristics of the designed GNRFETs with L=1.3μm, W=2.7μm. The minimum conduction point is 20.10 V.

Exact operation region and desired amount of hysteresis can be achieved by changing geometrical and material parameters of MLGNRFET. Figure 9 clearly hints that minimum gate voltage can be shifted nearly to zero voltage is also possible by geometrical and material engineering.

Beyond all these investigations, MLGNRFET posses many characteristics as a carbon family member such as high carrier mobility, superior heat conductance, availability and low cost. In addition, our MLGNRFET's very low dimension (L=1.3μm, W=2.7μm) is perfect for compact and dense integration. Moreover, our especially designed MLGNRFET is immune to leakage due to its geometrical structure. The above elaborated properties of MLGNRFET strongly prove its viability in emerging non-volatile market.

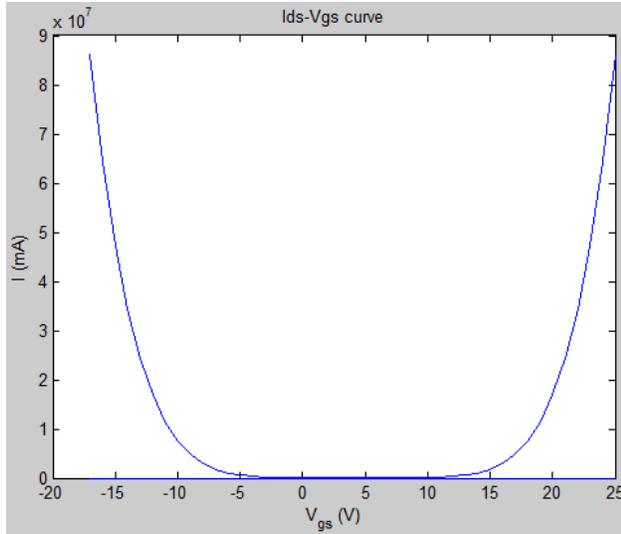

Figure 9: (non-ideal case) Ids–Vgs characteristics of the fabricated with L=2μm, W=2×75μm. The minimum conduction point is very close to 0 V.

## VII. CONCLUSION

In this work we have proposed ultrahigh dense, low power and high speed non-volatile memory devices by using a dual-gate, high $I_{on}/I_{off}$ ratio, high carrier mobility, high thermal conductive, hysteresis capable Graphene nanoribbon Field Effect Transistor (GNRFET). Carbon nanotube (CNT) based interconnect facilitate reduced contact resistance between with memory cell because they are from the same family. Also, Carbon nanotube (CNT) based interconnects consume less power with compared to the metallic counterpart like Cu and Al. Finally, the consolidate model will offer better non-volatile memory experience in terms of performance and reliability.